\documentclass[epj,twocolumn]{webofc}
\woctitle{CGS15}
\usepackage[varg]{txfonts}   
\usepackage[english]{babel}
\begin{document}
\selectlanguage{english}
\title{Disentangling the nuclear shape coexistence in even-even Hg
  isotopes using the interacting boson model} 

\author{J.E.\
  Garc\'{\i}a-Ramos\inst{1}\fnsep\thanks{\email{enrique.ramos@dfaie.uhu.es}} 
  \and 
  K. Heyde\inst{2}\fnsep\thanks{\email{kris.heyde@ugent.be}}
}

\institute
{Departamento de  F\'{\i}sica Aplicada, Universidad de Huelva,
21071 Huelva, Spain
\and
Department of Physics and Astronomy, Ghent University,
Proeftuinstraat, 86 B-9000 Gent, Belgium} 

\abstract{
We intend to provide a consistent description of the even-even Hg
isotopes, $^{172-200}$Hg, using the interacting boson model including
configuration mixing. We pay special attention to the description
of the shape of the nuclei and to its connection with the shape
coexistence phenomenon.  
}
\maketitle
\section{Introduction}
\label{intro}
Shape coexistence has been observed in many mass regions
throughout the nuclear chart and turns out to be realized in
more nuclei than anticipated a few decades ago \cite{heyde11}.  

Recently, a lot of new results have become available
for the even-even Po, Hg and Pt nuclei, for which experimental information was 
highly needed. 
In this mass region, the intruder bands are easily singled out for the
Pb and Hg nuclei  
and the excitation energies display the characteristic parabolic pattern with
minimal excitation energy around the $N=104$ neutron mid-shell
nucleus. In the case of Hg there is an intense experimental activity
for the light isotopes near the mid-shell region, which is
providing a very complete set of excitation energies, E2 transition
rates, isotopic shifts, $\alpha$-hindrance factors, etc. These new
results are painting a new landscape where the inclusion of intruder
states is a key ingredient to understand the physics of this mass
region. In particular, in a recent COULEX experiment for $^{184-188}$Hg
\cite{Bree14}, very detailed information on the shape of these nuclei
has been obtained.   

In a set of previous articles we studied the Pt \cite{Garc09,Garc11,Garc14b}
and the Hg \cite{Garc14a} nuclei extensively with the
Interacting Boson Model (IBM) \cite{Iach87}, incorporating
proton 2p--2h excitations (IBM-CM) \cite{duval82}. The conclusion of
these studies was that configuration mixing in the Pt nuclei is
somehow ``concealed'', while in the case of Hg its main effect does
not appear for the ground state but for the first two $2^+$ states.   

The IBM-CM 
allows the simultaneous treatment and mixing of several
boson configurations which correspond to different particle--hole
shell-model excitations \cite{duval82}. 
Hence, the model space  corresponds to a $[N]\oplus[N+2]$ 
boson space. The boson number $N$ is obtained as the sum of the number
of active protons  (counting both proton particles and holes) and the
number of valence neutrons, divided by two. 
Thus, the Hamiltonian for
two-configuration mixing is written as
\begin{equation}
  \hat{H}=\hat{P}^{\dag}_{N}\hat{H}^N_{\rm ecqf}\hat{P}_{N}+
  \hat{P}^{\dag}_{N+2}\left(\hat{H}^{N+2}_{\rm ecqf}+
    \Delta^{N+2}\right)\hat{P}_{N+2}\
  +\hat{V}_{\rm mix}^{N,N+2}~,
\label{eq:ibmhamiltonian}
\end{equation}
where $\hat{P}_{N}$ and $\hat{P}_{N+2}$ are projection operators onto
the $[N]$ and the $[N+2]$ boson spaces, 
respectively, $\hat{V}_{\rm mix}^{N,N+2}$  describes
the mixing between the $[N]$ and the $[N+2]$ boson subspaces, 
$\hat{H}^i_{\rm ecqf}$ is the extended consistent-Q Hamiltonian (ECQF)
with $i=N,N+2$ (see \cite{Iach87}), and $\Delta^{N+2}$ can be 
associated with the energy needed to excite two particles across the
$Z=82$ shell gap.

Within this formalism we have performed a fit to the excitation
energies and $B(E2)$ transition rates of $^{172-200}$Hg in order to
fix the parameters for the IBM-CM Hamiltonian. 
The results from the fitting procedure
are summarized in Table 3 of Ref.~\cite{Garc14a} and they will be used
in the calculations shown in the present contribution.

Our aim is 
to describe the shape evolution of Hg isotopes, paying special
attention to the mid-shell nuclei,
$^{180-188}$Hg. To do so, we present results i) from the
IBM-CM coherent state formalism, ii) for the quadrupole deformation
parameter $\beta$ extracted from the experimental $B(E2)$ values, and,
iii) on quadrupole shape invariants. 

\section{Shape evolution from different approaches}
\label{sec-comp}

The IBM-CM calculations,
carried out in Ref.~\cite{Garc14a}, display a strongly evolving character 
of the wave function in the $[N]$ and $[N+2]$ space along the Hg
isotope chain. The lightest and the heaviest Hg isotopes 
show a rather pure $[N]$ composition for the lowest lying states,
while the isotopes near the mid-shell, $^{180-186}$Hg, present a mixed
character (see Fig.~12 of Ref.\cite{Garc14a}),
especially for the lowest two $2^+$ states.  
These changes in the wave function are expected to
modulate the deformation of the nucleus.
\begin{figure}
\centering
\includegraphics[width=6cm,clip]{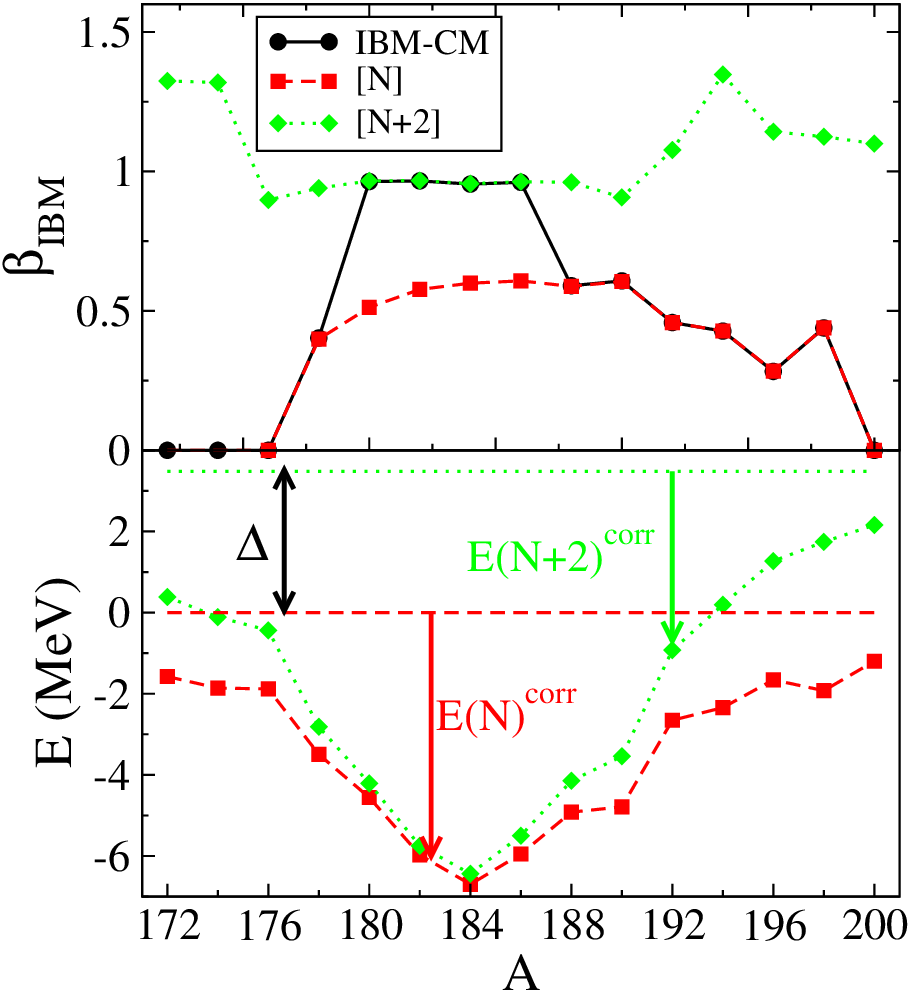}
\caption{Upper part: value of $\beta_{IBM}$ from the IBM-CM mean-field
  calculation. Lower part: 
  absolute energy of the lowest regular and lowest intruder states 
  for $^{172-200}$Hg. The arrows correspond to the correlation
  energies in the N and N+2 subspaces.}
\label{fig-beta}       
\end{figure}

A first approach to disentangle the shape evolution in Hg isotopes 
results from
the geometric interpretation of the IBM, which is obtained within the
intrinsic state formalism, proposed by Ginocchio {\it et
  al.}~\cite{Gino80}. 
In this approach one derives the equilibrium
value of the shape
variables $\beta_{IBM}$ (deformation) and $\gamma_{IBM}$ (degree of triaxiality)
by means of a variational procedure.
These variables can be connected with the standard Hill-Wheeler variables
\cite{Gino80,Garc14b}, in particular, $\beta_{IBM}$ is directly
connected with the deformation parameter of the collective model, 
$\beta$, while $\gamma_{IBM}=\gamma$. To study the 
geometry of the IBM-CM, Frank {\it et al.}~\cite{Frank06} proposed a
new method which takes into account the existence of two families of
states. 

We have calculated the energy surface for the Hg chain of isotopes and
the corresponding equilibrium value of the deformation parameter
$\beta_{IBM}$. In the upper part of Fig.~\ref{fig-beta} we present the value of
$\beta_{IBM}$ for the 
unperturbed configurations (omitting the mixing term in the
Hamiltonian), as the dashed line with squares, for the regular states, while
using a dotted line with diamonds for the intruder states. 
It is clear that the
unperturbed configurations result in a very different value of $\beta_{IBM}$
(around $0.5$ for the regular state and $1.0$ for the intruder one) and that
the full IBM-CM varies between the regular value at the beginning and
the end of the shell, and the intruder one at the mid-shell region. 

This can be partially
understood inspecting the  bottom part of Fig.~\ref{fig-beta}, where
the unperturbed energies for the lowest two $0^+$ states are shown. 
One observes very close-lying unperturbed configurations near
mid-shell, however, contrary to the situation in the Pt nuclei, the
intruder state never becomes the ground-state. 
Note that the full IBM-CM energy, though not shown, 
very closely follows the regular
unperturbed configuration in Fig.~\ref{fig-beta} (lower
part). Combining the upper (deformation) and the lower part (energies),
this looks like a {\it conumdrum}. As we shall explain later, this
results from the fact that energies obtained from mean-field
calculations do not include the full dynamics, which is particularly
important in the case of two close-lying configurations, as is the case
here. The
equilibrium value of $\gamma$ for the whole chain has also been 
calculated, giving rise to an oblate shape at the beginning and at the
end of the shell, while prolate around the mid-shell region, 
{\it i.e.}, $^{180-186}$Hg. 

\begin{figure}
\centering
\includegraphics[width=6cm,clip]{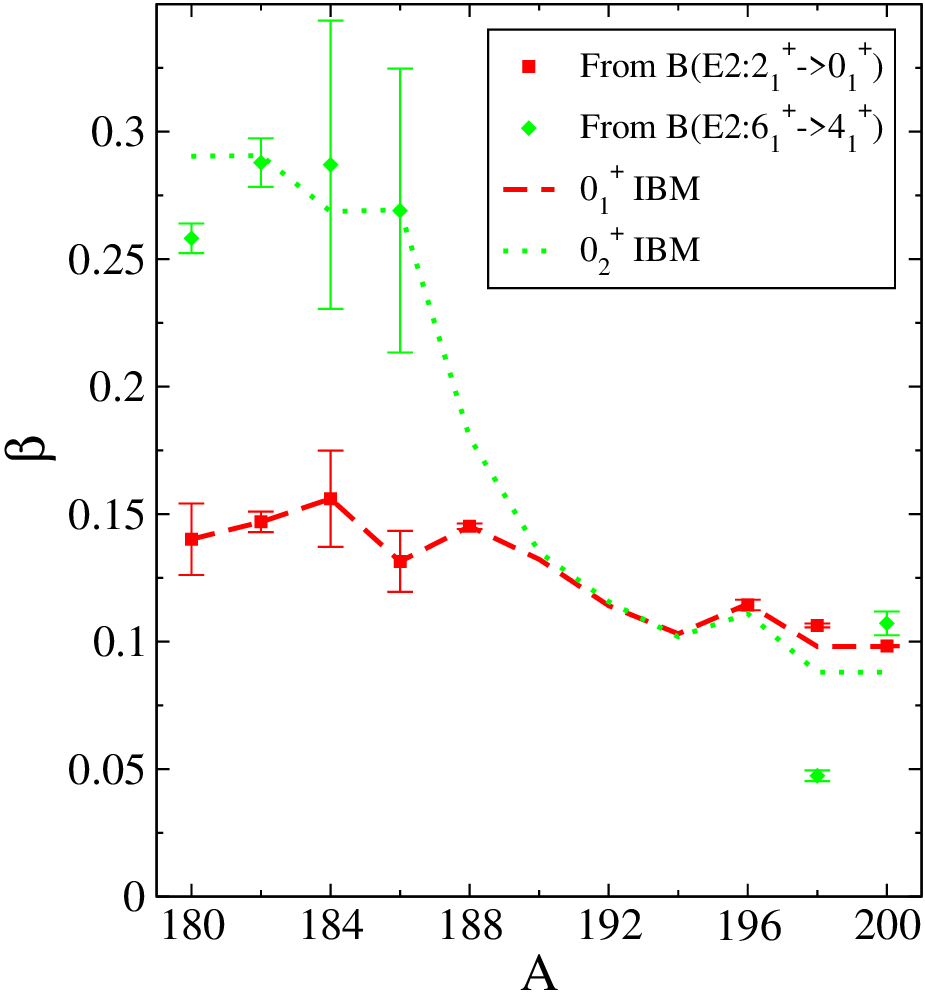}
\caption{Value of $\beta$ extracted from $B(E2)$ values.}
\label{fig-beta-be2}       
\end{figure}

In a second approach, we use a phenomenological interpretation   
to extract information on the quadrupole deformation $\beta$. 
The key point is to consider the geometrical view of the nucleus
and to extract the $\beta$ parameter
for a given band, either regular or intruder, from a known $B(E2)$ value
\begin{equation}
\beta=\frac{\sqrt{B(E2: I_i\rightarrow I_f)}}{\langle I_i K 2 0| I_f
  0\rangle \frac{3}{4\pi} Z e R_0^2},
\end{equation} 
where $R_0=1.2 A^{1/3}$ fm and $<.... | ..>$ is the Clebsch-Gordan
coefficient. 
Around the mid-shell region we assume that $0_1^+$ and
$2_1^+$ states belong to the regular band while the $6_1^+$ and
$4_1^+$ states belong to the
intruder band. In Fig.~\ref{fig-beta-be2}, we depict the value of
$\beta$ using this method. One can easily single out the presence of
two configurations with very different deformation; the lower
configuration, which corresponds to the regular state, is less deformed
than the higher one and can be identified with an intruder
state. This is an empirical evidence about the nature of the
ground state, which, close to mid-shell, always corresponds to
a less deformed and regular configuration, in agreement with
the lower panel of Fig.~\ref{fig-beta}, confirming the failure of the
IBM mean-field calculation in providing a correct value of
$\beta_{IBM}$. Therefore, the ground state is always regular, while
the $0_2^+$ band corresponds to an intruder configuration near
mid-shell, although the $\beta$ value drops dramatically from $A=188$
and onwards, denoting that the character of this band is no longer
intruder, but changes in character. 

In a third approach, we construct quadrupole invariants to
extract information about the nuclear deformation in a model
independent way \cite{kumar72,sreb11}.
Even though the shape of the nucleus is not an experimental
observable, it is still possible to extract from the data direct
information about various moments characterizing the nuclear shape
corresponding with a given eigenstate. Using Coulomb excitation, it is
possible to extract the most important diagonal and non-diagonal
quadrupole and octupole matrix elements, including their relative
signs and, in a model independent way, extract information about
nuclear deformation as shown by Kumar \cite{kumar72} (see also
\cite{sreb11}). 
\begin{figure}
\centering
\includegraphics[width=6cm,clip]{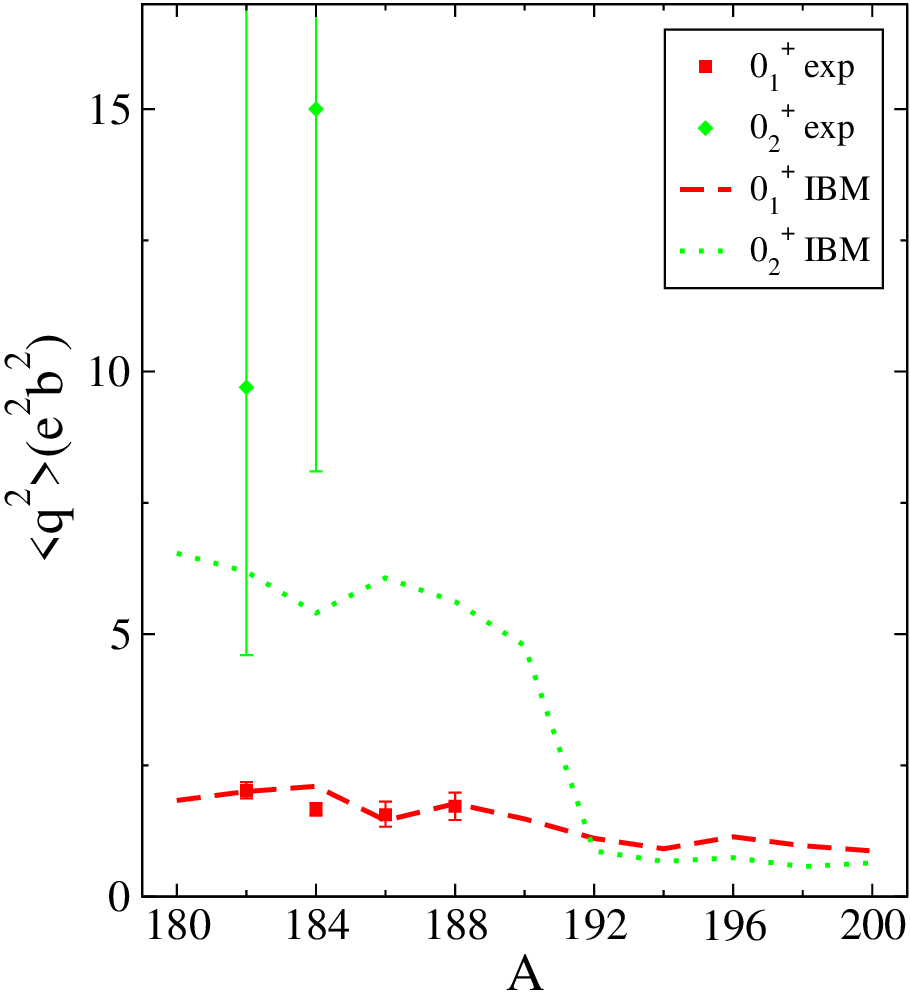}
\caption{Comparison of the experimental and theoretical $q^2$ shape
  invariant given in $e^2 b^2$ units for the first two $0^+$ states.}
\label{fig-3}
\end{figure}

From the theoretical point of view the nuclear shape can be calculated
using quadrupole shape invariants. In particular $q^2$ corresponds to,
\begin{equation} 
q^2=\sqrt{5} \langle 0^+| [\hat{Q} \times \hat{Q} ]^{(0)}|0^+ \rangle
=\sum_r \langle 0^+|| \hat{Q}|| 2_r^+\rangle  \langle
2_r^+||\hat{Q} ||0^+ \rangle, 
\label{q_invariant1}
\end{equation} 
where $q$ denotes the nuclear intrinsic quadrupole moment and the sum
is running over the complete basis of intermediate states with $L=2$. 

A comparison with the experimental values can be carried out whenever
a large enough set of reduced E2 matrix elements can  
be extracted from, \textit{e.g.}, Coulomb excitation experiments. Such
a comparison constitutes a very 
stringent test for the theoretical model and, at the same time, provides
a clear picture of the nuclear shape. 

In Fig.~\ref{fig-3} we compare the IBM-CM results with recent
Coulomb excitation experiments of $^{182-188}$Hg at REX-ISOLDE and
Miniball, allowing to extract a useful set of reduced E2 matrix
elements \cite{Bree14}. It turns out that our IBM-CM calculations
indeed give rise to values of 
$q^2$ 
that differ by a
factor of $\approx$ 3 between the $0^+_1$ and $0^+_2$ states around
the mid-shell region. The
general picture provided through this approach corresponds to a ground
state slightly deformed with regular character, while the $0_2^+$
state presents a rapid evolution being intruder around mid-shell region,
but turning into a regular character from $^{190}$Hg and onwards. 

\section{Conclusions} 
We have studied the shape evolution of a chain of Hg isotopes,
$^{172-200}$Hg, using the IBM-CM intrinsic state formalism. 
We also obtained values of $\beta$
extracted from  reduced E2 transition rates, and, finally we
considered quadrupole shape invariants. The mean-field approach 
describes very well the presence of
two different structures with very different deformations, but,
is not able to describe correctly the evolution of the
ground state deformation. The second approach and the quadrupole shape
invariants indicate compelling evidence that the ground state is
always slightly deformed while 
the structure  of the $0_2^+$ state changes dramatically along the
chain of Hg isotopes.

\section{Acknowledgment} 
Financial support from the ``FWO-Vlaanderen'' (KH and JEGR) and the
InterUniversity Attraction Poles Programme - Belgian State - Federal
Office for Scientific, Technical and Cultural Affairs (IAP Grant No.
P7/12, is acknowledged.  This work has also been partially supported
by the Spanish MINECO and the FEDER under Project No.
FIS2011-28738-C02-02 and by Spanish Consolider-Ingenio 2010
(CPANCSD2007-00042).

\end{document}